\documentstyle{article}
\def\be{\begin{equation}}
\def\ee{\end{equation}}
\def\ba{\begin{eqnarray}}
\def\ea{\end{eqnarray}}
\newcommand\lsim{\mathrel{\rlap{\lower4pt\hbox{\hskip1pt$\sim$}}
    \raise1pt\hbox{$<$}}}
\newcommand\gsim{\mathrel{\rlap{\lower4pt\hbox{\hskip1pt$\sim$}}
    \raise1pt\hbox{$>$}}}
\begin{document}
  \begin{flushright}
  SUSSEX-AST-94/7-1\\
  astro-ph/9407060\\
  \end{flushright}
\begin{center}
\bigskip
\large
{\bf First Order Inflation in General Relativity\footnote
{Talk presented at ``The Birth of the Universe'' workshop, Rome, May 1994}\\}
\smallskip
David Wands \\
\normalsize
\smallskip
{\it  Astronomy Centre,  School of Mathematical \& Physical Sciences, \\
University of Sussex,  Brighton BN1 9QH. U.~K.}\\
\end{center}
\bigskip
\begin{abstract}
I give a general formulation of the constraints on models of inflation
ended by a first order phase transition arising from the requirement
that they do not produce too many large (observable) true vacuum voids
-- the `big bubble problem'.  It is shown that this constraint can be
satisfied by a simple model in Einstein gravity -- a variant of `hybrid'
or `false vacuum' inflation.
\end{abstract}
\bigskip
%

The idea that inflation could be ended by a first order phase
transition is often dismissed as too problematic following Guth's
original trouble in finding a graceful exit from the false vacuum
\cite{Guth81}. The scalar field, $\psi$, whose self-interaction potential
acts like a cosmological constant driving the expansion, is trapped in
a local potential minimum so cannot evolve classically. A fixed
tunnelling rate to the true vacuum in a de Sitter universe leads to a
time invariant state. One expects the phase transition to complete
either at once or not at all.

First order inflation was revived in the guise of extended inflation
\cite{extinf} which invoked an extended gravitational lagrangian.  In
Brans-Dicke gravity \cite{BD}, where Newton's constant is replaced by
a field $\Phi$, the expansion rate decreases with time as the Planck
mass, $m_{\rm Pl}$, grows during inflation allowing the phase transition to
eventually complete. However this too was shown to produce too many
large true vacuum voids in any Brans-Dicke model that was sufficiently
close to general relativity to obey other observational limits
\cite{Weinberg89,LW91} -- the `big bubble problem'.  Even when the
Brans-Dicke parameter $\omega$ is allowed to vary in time
\cite{LW92a} it is difficult to achieve an acceptable model as the
same variation of the Hubble rate required to produce a
non-scale-invariant bubble spectrum also produces a
non-scale-invariant conventional perturbation spectrum incompatible
with models of large-scale structure formation \cite{LidLyt93}.

My aim here is to show that these problems need not necessarily be
present if it is the changing {\it shape} of the potential that varies
during inflation rather than (or in addition to) the Hubble rate. Soon
after La and Steinhardt's work it was pointed out by both Linde
\cite{Linde90} and Adams \& Freese \cite{Ada+Fre91} that a graceful
exit from first order inflation can also be achieved within general
relativity\footnote {Indeed this clearly must be the case at some
level as La and Steinhardt's Brans-Dicke model is itself conformally
related to general relativity plus another scalar field.}.  All that
is required is the presence of a second scalar field which can
interact with $\psi$ and slow-rolls while the $\psi$ field is
trapped in the false vacuum.  It is not obvious whether these models
based on two interacting fields in general relativity should fair any
better than the extended gravity models in avoiding the `big bubble
problem'. Indeed, as far as I am aware, this issue has never been
seriously addressed before. Here I will quantify the big bubble
constraints on the model proposed originally by Linde \cite{Linde90}
and give a more general formulation of the big bubble constraint in
first order inflation. For a wide range of parameter space, models of
first order inflation in general relativity easily satisfy these
constraints while producing near-scale-invariant perturbation spectra.
These results are based on work presented in \cite{CLLSW94}.

\section*{The Model}

To give probably the simplest example, consider the first order
potential proposed originally by Linde \cite{Linde90}
and discussed more recently as a variant of `hybrid' \cite{Linde94} or
`false vacuum' \cite{CLLSW94} inflation.
\ba
V(\psi,\phi) & = & {\lambda \over 4} \left( \psi^4 + M^4 \right)
		- {\gamma \over 3} M \psi^3
		+ {\alpha \over 2} M^2 \psi^2 \nonumber \\
& & \quad + {\lambda' \over 2} \phi^2 \psi^2
	+ {1 \over 2} m^2 \phi^2 \; .
\label{eV}
\ea
$\lambda$, $\lambda'$, $\alpha$ and $\gamma$ are all dimensionless
coupling constants which we can take to be of order unity. Requiring
the energy density of the true vacuum (at $\phi=0$) to vanish one can
eliminate $\gamma$ in terms of $\alpha$, say.  I will consider the
case where the mass $M$ is very much greater than $m$ and $\psi$ will
rapidly roll down to $\psi=0$ while $\phi$ is still large.


The effective potential for the field $\psi$ at different values of the
field $\phi$ is then that drawn in Figure~1,
for $0<\alpha<\gamma^2/4\lambda$. The mass of the $\psi$ field in the
false vacuum ($\psi=0$) is
\be
M_{\psi}^2 (\phi) = \tilde{\alpha}(\phi)M^2 \equiv
 \alpha M^2 + \lambda' \phi^2 \; .
\ee
Clearly a second order transition to the true vacuum is not possible
when $\alpha>0$.

A useful parameter describing the `shape' of the first order potential
for $\psi$ is given by
\be
\delta(\phi) = \frac{9\lambda\tilde{\alpha}(\phi)}{\gamma^2} \; .
\ee
For $\delta>9/4$ the false vacuum at $\psi=0$ is the only minimum of
$V(\psi)$. As $\phi$ continues to roll down towards its minimum at
$\phi=0$, $\delta$ decreases, and for $\delta<2$ the potential of the
second minimum, the true vacuum, becomes lower than the false vacuum,
although they are still separated by the potential barrier.
Eventually, as the energy difference between the two minima increases,
if the nucleation rate becomes large enough, the phase transition can
complete.

\section*{Big bubble constraints}

The most important parameter in determining the dynamics of a cosmological
first order transition is the percolation parameter
\be
p = {\Gamma \over H^4} \; ,
\ee
where the Hubble rate $H\equiv\dot{a}/{a}$ and $\Gamma$ is the
physical nucleation rate (i.e.~the number of bubbles of the true
vacuum nucleated per unit physical volume per unit time).

To produce a successful model of first order inflation the percolation
parameter must fulfil two competing requirements. Firstly it must
remain small enough for long enough to produce a large, flat,
homogeneous, etc.\ universe solving the horizon, flatness, smoothness,
etc.\ problems with an acceptably small filling fraction of large true
vacuum voids nucleated early on during inflation and swept up by the
subsequent expansion ($p\lsim p_*$). But eventually the percolation
parameter must reach its critical value in order to complete the
transition to the true vacuum phase ($p\gsim p_{\rm cr}\sim 1$). As
remarked earlier, in models based on extended gravity theories this
may be achieved by allowing the effective gravitational ``constant''
to vary producing a decreasing Hubble rate and thus an increasing $p$.
The precise big bubble constraints have been considered in the case of
both extended \cite{LW91} and hyperextended \cite{LW92a} inflation.
Here I will approximate these results by the requirement that the
filling fraction of voids nucleated $e^{55}$ expansion times
($55$ e-foldings) from the end of inflation be less than
$10^{-5}$ ($p_{55}\lsim 10^{-5}$).

We can give these constraints in terms of the shape of the first order
potential using the calculation of the tunnelling rate given originally
by Coleman \cite{Coleman77}
\be
\Gamma = A \exp(-S_E) \; ,
\ee
with $A\sim M^4$ and $S_E$ is the Euclidean action of the tunnelling
configuration\footnote{This flat spacetime result is valid except in the
extreme thin wall limit which turns out not to be relevant for
cosmologically interesting scenarios \cite{LW92b}.}.
Recently Adams \cite{Adams93} has shown that this can be given
in terms of a numerically calculated fitting function of the shape of the
potential:
\be
S_E = {2\pi^2 \over \lambda} B_4 (\delta) \; .
\ee

Thus requiring $p\geq p_{\rm cr}$ to end inflation places an upper bound
on the minimum value of the parameter $\delta$ as $\phi\to0$
\be
\delta_0 \equiv {9\lambda\alpha \over \gamma^2} \; ,
\ee
such that
\be
B_4(\delta_0) < B_4(\delta_{\rm cr}) =
 {\lambda \over 2\pi^2} \ln {\lambda M^4 \over 4p_{\rm cr}H^4} \; .
\ee
If the transition is too strongly first order for a given energy scale
$M$, i.e.~$\delta_0>\delta_{\rm cr}(M)$, the transition can never
complete and inflation continues forever.

On the other hand requiring
$p_{55}\lsim 10^{-5}$ corresponds to an lower limit
on the parameter $\delta$ $55$ e-foldings from the end of inflation.
\be
B_4(\delta_{55}) \gsim B_4(\delta_*) =
 {\lambda \over 2\pi^2}
	\left( \ln {\lambda M^4 \over 4p_{\rm cr}H^4} + 11.5 \right) \; ,
\ee
where the calculated values of $\delta_{\rm cr}$ and $\delta_{*}$ are plotted
in Figure~2.
Models such as extended inflation which only alter the mass scales must
proceed along a horizontal trajectory (i.e.\ at fixed $\delta$) from
outside $\delta_{55}$ to reach $\delta_{\rm cr}$ during the
last $55$ e-foldings of inflation, whereas if the shape of the potential
changes (changing $\delta$) they can also proceed vertically.


\section*{Conclusions}

Returning to our specific model, $\delta_{\rm cr}$ fixes the value of
$\phi$ at the end of inflation, and we also have a minimum value for
$\delta$ and thus $\phi$ $55$ e-foldings earlier. In the limit where
$\lambda M^4\gg m^2\phi^2$ the potential energy density remains
approximately constant during inflation and the ratio between the
value of $\phi$ at the end of inflation and the value $N$ e-foldings
earlier is $\exp(Nm^2m_{\rm Pl}^2/2\pi\lambda M^4)$ so we can
translate this big bubble constraint into a bound on the mass scales
\cite{CLLSW94}
\be
\frac{m^2m_{\rm Pl}^2}{M^4} \gsim
 {\pi\lambda\over 55}
 \ln \left( {\delta_*-\delta_0 \over \delta_{55}-\delta_0} \right) \; .
\ee
As the expression on the right-hand-side is likely to be $\lsim
10^{-2}$ for $\lambda\lsim 1$ (as long as $\delta_{\rm cr}$ is not too
close to $\delta_0$) this bound does not threaten to force us out of
the extreme slow-rolling limit. If we demand that it is fluctuations
in the $\phi$ field that are the near-scale-invariant seed density
perturbations for large-scale structure\footnote {We can obtain small
quantum fluctuations in the slow-rolling $\phi$ field in this model
without introducing small coupling parameters due to the disparate
mass scales $m$ and $M$.} then this gives us another relation between
the mass scales \cite{CLLSW94} turning the big bubble constraint into
a lower bound on both $m$ and $M$ individually.  For values of the
coupling constants of order unity this constrains $M$, say, to be
greater than about $10^{13}$GeV.

Thus even while the Hubble rate stays very nearly constant (and the
perturbation spectrum is very nearly flat \cite{CLLSW94}) it is
possible to ensure a sufficiently rapid growth in the percolation
parameter after observable scales start leaving the horizon and bring
inflation to a graceful end.  Any decrease in the Hubble parameter
only accelerates this change in the percolation parameter.

Models which rely solely on a decreasing Hubble rate to increase the
percolation parameter tend to run into problems as this also changes
the tilt of the conventional perturbation spectrum \cite{LidLyt93}.
Any models based on alternative gravity theories which also change the
potential shape may also be viable \cite{LayLid94,Linde94} or even
phenomenologically interesting \cite{AmeOcc94}, but invoking an extended
gravity theory is not necessary in order to successfully end inflation
by a first order phase transition.

\smallskip
I am grateful to my colleagues for allowing me to present here work from
our joint paper \cite{CLLSW94}. The author is supported by the PPARC and
acknowledges use of the Starlink computer system at Sussex.



\section*{Figure captions}

\noindent
{\em Figure 1.} $V(\psi)$ given in Eq.~(\ref{eV}) for
$\alpha=\lambda=1$ at five different values of $\phi$ and thus
different $\delta$.

\bigskip
\noindent
{\em Figure 2.} $\delta_{\rm cr}$ and $\delta_*$ plotted against
$\log_{10}(V^{1/4}/m_{\rm Pl})$


\frenchspacing
\begin{thebibliography}{99}
\parskip0pt
\bibitem{Guth81} A. H. Guth, {\it Phys. Rev. D}{\bf 23} 347 (1981)
\bibitem{extinf} D. La \& P. J. Steinhardt, {\it Phys. Rev. Lett.} {\bf 62} 376
 (1989)
\bibitem{BD} C. H. Brans \& R. H. Dicke, {\it Phys. Rev.} {\bf
 124} 925 (1961)
\bibitem{Weinberg89} E. J. Weinberg, {\it Phys. Rev. D}{\bf 40} 3950 (1989)
\bibitem{LW91} A. R. Liddle \& D. Wands, {\it Mon. Not. Roy. astr. Soc.}
 {\bf 253} 637 (1991)
\bibitem{LW92a} A. R. Liddle \& D. Wands, {\it Phys. Rev. D}{\bf 45} 2665
 (1992)
\bibitem{LidLyt93} A. R. Liddle \& D. H. Lyth, {\it Phys. Rep.} {\bf 231} 1
 (1993)
\bibitem{Linde90} A. D. Linde, {\it Phys. Lett.} {\bf B249} 18 (1990)
\bibitem{Ada+Fre91} F. C. Adams \& K. Freese, {\it Phys. Rev. D}{\bf 43} 353
 (1991)
\bibitem{CLLSW94} E. J. Copeland, A. R. Liddle, D. H. Lyth, E. D. Stewart
 \& D. Wands, {\it Phys. Rev. D}{\bf 49} 6410 (1994)
\bibitem{Linde94} A. D. Linde, {\it Phys. Rev. D}{\bf 49} 748 (1994)
\bibitem{Coleman77} S. Coleman, {\it Phys. Rev. D}{\bf 15} 2929 (1977)
\bibitem{LW92b} A. R. Liddle \& D. Wands, {\it Phys. Rev. D}{\bf 46} 3655
 (1992)
\bibitem{Adams93} F. C. Adams, {\it Phys. Rev. D}{\bf 48} 2800 (1993)
\bibitem{LayLid94} A. M. Laycock \& A. R. Liddle, {\it Phys. Rev. D}{\bf 49}
 1827 (1994)
\bibitem{AmeOcc94} L. Amendola, S. Capozziello, M. Litterio \& F. Occhionero,
 {\it Phys. Rev. D}{\bf 45} 417 (1992)
\end{thebibliography}
\end{document}